\documentclass[pre,twocolumn]{revtex4-1}
\usepackage{subcaption}
\usepackage{graphicx}
\usepackage{comment}
\usepackage{amsmath,amssymb,amsthm} 
\usepackage{bm}
\usepackage{verbatim}
\usepackage{float}
\begin{document}

		\title{ Discrete Molecular Dynamics }

\author{ S\o ren  Toxvaerd }
\affiliation{ Department
 of Science and Environment, Roskilde University, Postbox 260, DK-4000 Roskilde, Denmark}
\date{\today}

\vspace*{0.7cm}

\begin{abstract}

Computer simulation of the time evolution in a classical system is a standard numerical method,
	used in numerous scientific articles in Natural Science. Almost all the simulations are performed by discrete Molecular Dynamics (MD).
	The algorithm used in MD was originally
	formulated by  I. Newton at the beginning of his book $Principia$.
Newton's discrete dynamics is exact in the same sense as Newton's analytic counterpart Classical Mechanics. Both dynamics are
 time-reversible, symplectic, and have the same dynamic invariances.
There is no qualitative difference between the two kinds of dynamics. This is due to the fact, that there exists a ''shadow Hamiltonian''
	nearby the Hamiltonian $H(\textbf{q},\textbf{p})$ for the analytic dynamics, and where its dynamics can be obtained by an asymptotic expansion
	from $H(\textbf{q},\textbf{p})$, and where the positions generated by MD are located on the analytic trajectories for the shadow Hamiltonian.

It is only possible to obtain the solution of  Newton's classical differential equations for a few  simple systems, but
the exact discrete Newtonian dynamics can be obtained for almost all complex classical systems.
Some examples are given here: The emergence and evolution of a planetary system. The emergence and evolution of planetary systems with inverse forces. 
The emergence and evolution of galaxies in the expanding Universe.

The fact that there exist two equally valid formulations of classical dynamics raises the question:
What is the classical limit of quantum mechanics? Discrete molecular dynamics is mathematically different from analytic dynamics.
	The Heisenberg uncertainty between the concurrent values
	of positions and momenta is an inherent property of the discrete dynamics, but
 the analytic quantum electrodynamics (QED) is in all manner fully appropriate, and there is a lack of justification for preferring discrete quantum mechanics.
\end{abstract}
\maketitle

\section{ Introduction}
 In Molecular Dynamics (MD) the movements of atoms and molecules are obtained by Newtonian dynamics \cite{Newton1687}.
 The trajectories of atoms and molecules are determined by numerically
solving Newton's equations of motion for a system of interacting particles, where forces between the particles and
their potential energies are calculated using interatomic potentials or molecular mechanics force fields.
The method is widely applied in physics, chemistry, materials science, biophysics, and biochemistry.
 
 Almost all MD simulations are performed using a simple discrete algorithm, mostly named the \textit{Verlet algorithm} (Appendix A) \cite{Verlet1967}. But 
 the algorithm also appears under a variety of other names e.g. Leap frog, and in textbooks for MD \cite{Tildesley,Frenkel} the 
 algorithm is presented as a 
 third-order predictor of the positions of the objects in the system. It was, however, 
  Isaac Newton who first formulated the \textit{Discrete Molecular Dynamics} algorithm, when he in
  PHILOSOPHI\AE \ NATURALIS PRINCIPIA MATHEMATICA.  $(Principia)$ \cite{Newton1687}
 derived his second law for classical mechanics. The discrete algorithm is not only time-reversible and
 symplectic, but it has also the same dynamic invariances for a conservative system (momentum, angular momentum, and energy)
 as Newton's analytic dynamics. So the dynamics obtained by the algorithm is exact in the
 same sense as an exact solution for Newton's analytic Classical Mechanics. Furthermore, the dynamics obtained by Newton's analytic classical mechanics and
 his discrete molecular dynamics are qualitatively equal, because there exists
 a \textit{shadow Hamiltonian} nearby the  Hamiltonian for the analytic dynamics, and for which the discrete positions are placed on
 the analytic trajectories for the shadow Hamiltonian \cite{toxone}.

 In the chapter on Discrete Molecular Dynamics we first in Section 2  present Newton's formulation of the analytic classical dynamics and the discrete dynamics
  in $Principia$.
 The proof that the two complementary formulations of classical dynamics, the analytic dynamics, and the discrete dynamics
 have the same dynamical invariances, is given in Section 3.1. The connection between the two kinds of dynamics,
 given by the existence of a shadow Hamiltonian,  
 is given in Section 3.2. The mathematical difference between the two formulations and the connections with discrete quantum mechanics is presented in Section 3.3.
  In Section 4 there are three examples of exact discrete molecular dynamics for complex systems.
The examples are 4.1. The emergence and evolution of a planetary system. $4.2.$ The emergence and evolution of planetary systems with inverse forces. 
4.3. The emergence and evolution of galaxies in the expanding Universe. The article ends with a conclusion in Section 5.

 \section{Newtonian dynamics}
Isaac Newton  (1643-1727) published PHILOSOPHI\AE \ NATURALIS PRINCIPIA MATHEMATICA  $(Principia)$ \cite{Newton1687} in 1687,
where he formulated the equations for the classical analytic dynamics of objects. 
$Principia$ was reprinted in 1713 with errors of the 1687 edition corrected, and in an improved version in 1726.

\subsection{$Principia$}
The first edition of \textit{Principia} contains  510 pages, many with figures and geometrical proofs of the
dynamics of a celestial object.
\textit{Principia} starts with  \textit{DEFINITIONES}, where Newton defines \textit{matter} (mass) 
 \textit{motion} (momentum) and \textit{force}, and in the next section Newton's  three laws are postulated in
 \textit{AXIOMATA sive LEGES MOTUS}  on page 12-25, and with detailed explanations in \textit{Corollary I-VI}.
The English translation \cite{Newtonengtrans} of the Latin formulation of Newton's second law is

\textit{ The alteration of motion}(momentum) \textit{is
ever proportional to the motive force impressed; and is made in the direction of the right line in which that force
is impressed.}\\
Expressed with Newton and Leibniz's   algebra the second law for the differential change of the  momentum $\textbf{p}$ of a mass center due to a 
force $\textbf{F}$ is 

\begin{equation}
	\frac{d \textbf{p}(t)}{dt}= \textbf{F}(\textbf{r}(t))
\end{equation}

After the section with the axioms of his three laws and the corollaries  follows   a section on pages 26 - 36, 
\textit{De MOTU CORPORUM} ( OF THE MOTION OF BODIES),
where Newton in eleven lemmas treats the connections between infinitesimals and the limiting procedure for analytic curves.

Newton's second law for classical analytic mechanics as well as
his discrete dynamics are presented in the succeeding section  in  $\textit{Proposition I}$ at page 37 and  $\textit{Proposition VI}$ at page 44.

\subsection{$Proposition$ $I$ and $Proposition$ $VI$}

Newton's classical mechanics with the two $Propositions$ starts with the heading
\textit{Of the Invention of Centripetal Forces.}
\textit{Proposition I}  is  a derivation/illustration of Newton's second law and how to obtain the  analytic  orbit of a mass center
 by Newton's discrete dynamics.
 \textit{Proposition I}  is the central point
 in $Principia$, although it is presented as a geometrical proof of some equalities between areas in his discrete dynamics.
 The text was illustrated by a figure (Figure 1).  Newton supplemented the proof  in a succeeding proposition, \textit{Proposition VI},
 by considering the 
 change of the position of a mass center on an analytic orbit caused by an analytic centripetal force. 

  \begin{figure}
	  \begin{center}	  
 	 \includegraphics[width=8.6cm,angle=0]{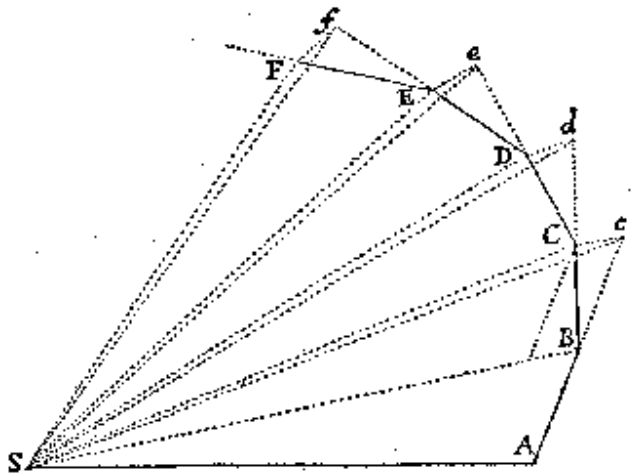}
 	 \caption{  Newton's figure at $Proposition$ $I$ in Principia, with the formulation of the discrete dynamics. 
	  The discrete positions are A: $\bf{r}$$(t-\delta t)$;  B: $\textbf{r}(t)$;  C: $\textbf{r}(t+\delta t)$, etc.. The deviation cC
	  from the straight line, ABc (Newton's first law) is caused by a force from the position S at time $t$.}
	  \end{center}		  
  \end{figure}
 The English translation  of  $\textit{Proposition I}$  is:\\
PROPOSITION I. Theorem I.\\
\textit{The areas, which revolving bodies describe by radii drawn to an immovable centre of force do lie in the same immovable planes, and
 are proportional to the times in which they are described}.\\
\textit{ For suppose the time to be divided into equal parts, and in the first part of time let the body by its innate force describe the right line
AB. In the second part of that time, the same would (by Law I.), if not hindered, proceed directly to c, along the line Bc equal to AB; so that the radii
AS, BS, cS, drawn to the centre, the equal areas ASB, BSc, would be described. But when the body is arrived at B, suppose that a centripetal force acts at once
with a great impulse, and, turning aside the body from the right line Bc, compels it afterwards
to continue its motion  along the right line BC. Draw cC parallel
to BS meeting BC in C; and at the end of  the second part of the time, the body (by Cor. I of Laws) will be found in C, in the same plane with the
triangle ASB. Join SC, and,  because SB and Cs are parallel, the triangle SBC will be equal to the triangle SBc, and therefore also to the
triangle SAB. By the like argument, if the centripetal force acts successively in C, D, E, \& c., and makes the body,
in each single particle of time, to describe the right lines CD, DE, EF, \& c., they will all lie in the same plane;
and the triangle SCD will be equal to the triangle SBC, and SDE to SCD, and SEF to SDE. And therefore, in equal times, equal areas are described in on immovable plane:
and, by composition, any sums SADS, SAFS, of those areas, are one to the other as the times in which they are described. Now let the number of
those triangles be augmented; and their breadth diminished in infinitum; and (by Cor. 4, Lem III) their ultimate perimeter ADF will be a curve line:
and therefore the centripetal force, by which the body is perpetually drawn back from the tangent of this curve, will act continually; and any described
areas SADS, SAFS, which are always proportional to the times of description, will, in this case also, be proportional to those times.} Q. E. D.

The text can be expressed with Newton's and Leibniz's algebra. The time is changed with a discrete and constant time increment $\delta t$.
The  body with mass $m$ is at the position A: $\textbf{r}(t-\delta t)$ at time $t-\delta t$ , at position  B: $\textbf{r}(t)$   at  time $\textit{t}$   and  
 position  C: $\textbf{r}(t+\delta t)$   at time $t +\delta t$.
The  momenta $ \textbf{p}(t+\delta t/2) =  m(\textbf{r}(t+\delta t)-\textbf{r}(t))/\delta t$  and
 $  \textbf{p}(t-\delta t/2)=  m(\textbf{r}(t)-\textbf{r}(t-\delta t))/\delta t$ are constant in
the time intervals in between the discrete positions. 
 In vector notation $\textit{Proposition I}$ is 
\begin{equation}
	\overrightarrow{BC}= \overrightarrow{Bc}+\overrightarrow{cC},
\end{equation}	
and the  momentum  is changed at $B$ by (Figure 1)
\begin{equation}
	 m\frac{\textbf{r}(t+\delta t)-\textbf{r}(t)}{\delta t}
					=m\frac{\textbf{r}(t)-\textbf{r}(t-\delta t)}{\delta t} +\delta t \textbf{f}(t).	
					 \end{equation}
One obtains the Verlet algorithm by a rearrangement of  Eq. (3) (see Appendix A)
\begin{equation}
	\textbf{r}(t+\delta t)= 2\textbf{r}(t)-\textbf{r}(t-\delta t) +\frac{\delta t^2}{m} \textbf{f}(t).	
					 \end{equation}
The corresponding ''Leap-frog" algorithm is
\begin{eqnarray}
  \textbf{p}(t+\delta t/2)= \textbf{p}(t-\delta t/2)+ \delta t \textbf{f}(t) \nonumber \\
	\textbf{r}(t+\delta t)= \textbf{r}(t)+ \frac{\delta t}{m}  \textbf{p}(t+\delta t/2).
					 \end{eqnarray}

 Newton notices that the relation, Eq. (3) for the dynamics holds for
any value of $\delta t$, and the differential equation for the change in momentum of a mass $m$ on a classical analytic orbit (Eq. (1)) is obtained as 
 the limit $ lim_{\delta t \rightarrow 0}$ of Eq. (3).

 Newton published the two new editions of $Principia$ with corrections and additions,
 but he never changed anything in the formulation of $\textit{Proposition I}$.   There
 are, however, several things to note about the formulation:

 1. The equality of the areas of the triangles is a consequence of the conservation of the angular momentum \cite{Toxvaerd2020}, and Newton was
 aware of this fact.  In astronomy, it is Kepler's second law for the Solar system,
 but Newton did not mention this important fact in $\textit{Proposition I}$.

 2. The equality of the areas plays no role in obtaining the analytic dynamics.

 3. The formulation is for discrete dynamics. But the discreteness is not only in time, space, and momentum, but also the force acts discrete:
\textit{..suppose that a centripetal force acts at once
with a great impulse,...}(Notice \textit{ at once}).

 When Newton formulated the second law, he must immediately have realized that his discrete relation
Eq. (3) at least explains Kepler’s second law. But he 
   did not mention it at the derivation of the second law, nor in his  second- or third editions of $Principia$.  An explanation for this omission could be, that
      Newton believed that the exact classical dynamics first is achieved in the analytic limit with continuous time and space. But this is in fact not the
       case, his discrete dynamics have the same invariances as his analytic dynamics (see  Section 3).

Newton returned in \textit{Proposition VI} to the determination of the analytic orbit of an object exposed to a centripetal force.
The figure at $Proposition$  $VI$ in $Principia$ (Figure 2) shows the analytic orbit ($APQ$) of a mass center continuously attracted
by a centripetal force from a force center at $S$. Newton proved that a tangential deviation  $ \overrightarrow{RQ}$ from the analytic
orbit after a time increment $\delta t$ is proportional to $\delta t^2$ and thus vanishes in
the limit $ lim_{\delta t \rightarrow 0}$  \cite{Chandrasekbar}. 

 \textit{Proposition I}  is the central point
 in $Principia$, although it is presented as a geometrical proof of some equalities between areas in discrete dynamics.
  \textit{Proposition I} and the relation with \textit{Proposition VI}  have played a crucial role in the history of
  $Principia$ and classical mechanics.
  The prehistory and genesis of  $Principia$ is given in a recent and excellent review article by Michael Nauenberg \cite{Nauenberg2019}. 
  The present article concerns the actual difference between the two formulations of classical dynamics.

In Summary:  \textit{Proposition I} is Newton's discrete dynamics and  \textit{Proposition VI} is for Newton's
Classical Mechanics. As will be shown in the next section both dynamics are exact and 
the two dynamics are qualitatively equal despite the fact that they mathematically are different.

  \begin{figure}
	  \begin{center}	  
 	 \includegraphics[width=8.6cm,angle=0]{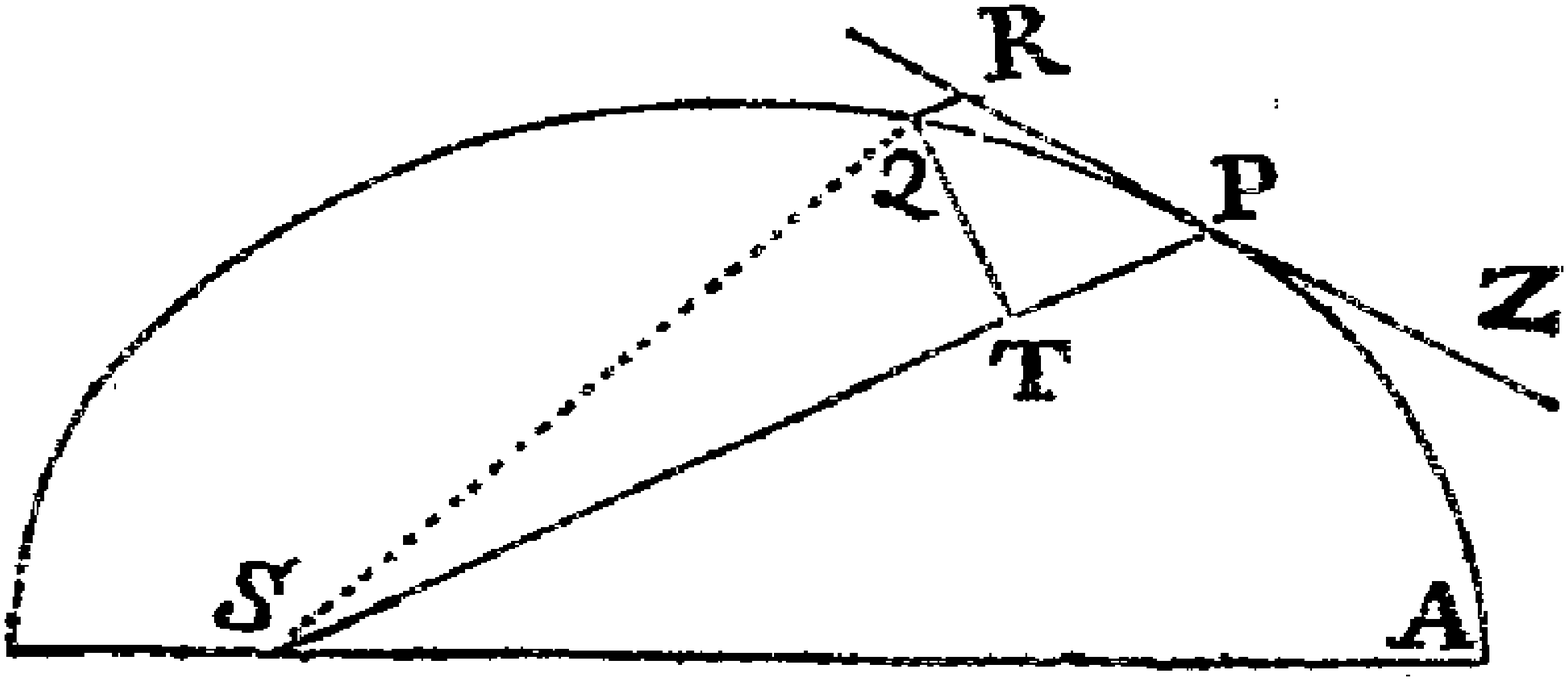}
	  \caption{  Figure in $Principia$ illustrating $\textit{Proposition VI}$.}
	  \end{center}		  
  \end{figure}
 $ $\\

 \subsection{The emergence and history of $Proposition$ $I$ and $Proposition$ $VI$ and Newton's second law} 

 Newton submitted in 1684-1685 a draft, $\textit{De Mutu Corporum Gyrum}$,  of $Principia$ to Edmond Halley and the Royal Society,
 and he and Robert Hooke  used in the period 1679-1684 the geometrical construction of discrete dynamics to obtain the orbits of 
 mass centers exposed to a discrete centripetal force \cite{Nauenberg2019}. The relationship between Newton and Hooke and the emergence
 of  $\textit{Proposition I }$  and Newton's second law have given rise to an extensive debate
 right up to our time \cite{Erlichson1997,Nauenberg1998,Nauenberg2005,Pourciau2011,Nauenberg2012,Pourciau2012}.
 Robert Hooke has without any doubt inspired Newton according to M. Nauenberg, although Newton not
 acknowledged Hooke's role in the formulation of
   $\textit{Proposition I }$  and Newton's second law.

   $\textit{Proposition I }$ is the algorithm for Discrete Molecular Dynamics, whereas  $\textit{Proposition VI }$  expresses the
   first-order deviation of the position by a discrete move from the analytic orbit. The equivalence of  $\textit{Proposition I }$ and $\textit{Proposition VI}$ was debated already shortly after
  the publication of $Principia$ among leading scientists (e.g. Leibniz and Huygens)\cite{Guicciardini1999,Nauenberg2014}.

  Newton's second law is traditionally not given by Eq. (1), but as an equality between the acceleration $\textbf{a}$ and force
\begin{equation}
	\textbf{F}(t)=m \textbf{a}(t)= m \frac{d^{2}\textbf{r}(t)}{d t^{2}}.
\end{equation}

 This algebraic formulation of Newton's second law was, however, first given after Newton's dead 1727 by Euler in 1736 \cite{Coelho2018}.

 \section{Newton's discrete and analytic dynamics} 

 Newton's Classical Mechanics is the exact formulation of the classical analytic dynamics of objects exposed to forces.
 The dynamics are obtained by solving the second-order 
 differential equations Eq. (6) for $N$  objects. The ''exactness'' is given by
 some qualities: the analytic dynamics is time reversible, symplectic, and with three invariances for a conservative system of $N$ objects. The
 total momentum, angular momentum, and energy of a conservative system are conserved. Here we shall first show (Subsection 3.1),
 that Newton's discrete dynamics, obtained by solving the
 corresponding discrete equations, Eq. (3), has the same quality and thus is exact in the same sense as his analytic classical mechanics.

 The analytic and the discrete dynamics is, however, most likely connected  by the existence of a
   $\textit{shadow Hamiltonian}$ for analytic dynamics \cite{toxone}, where the positions obtained by discrete dynamics are on the trajectories for the shadow Hamiltonian.
The indication of an existence of a shadow Hamiltonian is given in Subsection 3.2. An existence implies that the classical dynamics obtained by analytic and discrete dynamics, respectively,
are qualitatively similar. But the discrete dynamics and the analytic dynamics are, however, mathematically different and fundamentally with
different physics.  The difference between the two kinds of dynamics and the relation with quantum mechanics 
 is given in Subsection 3.3.

  \subsection{Newton's discrete dynamics}

  Newton's discrete dynamics for a simple system of  $N$ spherically symmetrical objects
     with masses $ m^N \equiv m_1, m_2,..,m_i,..,m_N$ and positions   \textbf{r}$^N(t) \equiv$ \textbf{r}$_1(t)$, \textbf{r}$_2(t)
     ,..,$\textbf{r}$_i(t),..$\textbf{r}$_N(t)$  is obtained 
 by Eq. (3). Let the force, $ \textbf{F}_i$ on object No $i$ be a sum of pairwise  forces  $ \textbf{f}_{ij}$ between pairs of   objects $i$ and $j$
 \begin{equation}
	 \textbf{F}_i=  \sum_{j \neq i}^{N} \textbf{f}_{ij}.
 \end{equation}	

   Newton's discrete dynamics, Eq. (3) is a central difference algorithm and it is time symmetrical, so the discrete dynamics is time reversible and symplectic \cite{Friedman1991}.
  
 The momentum for a conservative system of the $N$ objects is conserved since (Eq. (3))
\begin{eqnarray}
	\sum_i^N m_i\frac{\textbf{r}_i(t+\delta t/2)-\textbf{r}_i(t)}{\delta t}= 
\sum_i^N \textbf{p}_i(t+\delta t/2)= \\ \nonumber
\sum_i^N \textbf{p}_i(t-\delta t/2)+ \delta t\sum_{i,j \neq i}^N \textbf{f}_{ij}(t)=
\sum_i^N \textbf{p}_i(t-\delta t/2).
\end{eqnarray}

( $\sum_{i,j \neq i}^N \textbf{f}_{ij}(t)=0$ with $ \textbf{f}_{ij}(t)=-\textbf{f}_{ji}(t)$ due to Newton's third law). 

The discrete positions and momenta are not
 known simultaneously. An expression for the angular momentum of the conservative system is 
\begin{eqnarray}
	\textbf{L}(t)=\sum_i^N \textbf{r}_i(t)  \times (\textbf{p}_i(t+\delta t/2) +\textbf{p}_i(t-\delta t/2) )/2 \nonumber \\
	= \sum_i^N \textbf{r}_i(t)  \times  (m_i \textbf{r}_i(t+\delta t) -m_i \textbf{r}_i(t-\delta t) )/ 2\delta t   ).
\end{eqnarray}	
The angular momentum is conserved since (using $\textbf{r}_i(t)  \times ( \textbf{f}_{ij}(t)+\textbf{f}_{ji}(t)) =0,$
  $\textbf{a} \times \textbf{a}=0,$ $ \textbf{a} \times \textbf{b}=-  \textbf{b} \times \textbf{a}$ and  Eq. (4)  )
 	\begin{center}	 
\begin{eqnarray} 
 2 \delta t \textbf{L}(t)=\sum_i^N \textbf{r}_i(t) \times (m_i \textbf{r}_i(t+\delta t) -m_i \textbf{r}_i(t-\delta t) )  \nonumber \\
=\sum_i^N m_i \textbf{r}_i(t) \times (2\textbf{r}_i(t) -2\textbf{r}_i(t-\delta t) ) \nonumber \\ 
=\sum_i^N m_i\textbf{r}_i(t-\delta t) \times (\textbf{r}_i(t) +\textbf{r}_i(t))  \nonumber \\
=\sum_i^N m_i\textbf{r}_i(t-\delta t) \times (\textbf{r}_i(t) -\textbf{r}_i(t-2\delta t)) \nonumber \\
=2 \delta t \textbf{L}(t-\delta t).
\end{eqnarray}	
\end{center}

 The  energy in analytic dynamics is the sum of potential energy $ U(\textbf{r}^N(t))$ and kinetic energy $K(t)$, 
 and it is an invariance for a conservative system.
 The kinetic energy in the discrete dynamics is, however, ill-defined since the velocities at time $t$ are not known. Traditionally
one uses the expressions
\begin{eqnarray}
 \textbf{v}_i(t)=\frac{\textbf{r}_i(t+\delta t) -\textbf{r}_i(t-\delta t)}{2\delta t} \\
 K_0(t)=  \sum_i^N \frac{1}{2}m_i \textbf{v}_i(t)^2 \\
 E_0= U(\textbf{r}^N(t))	+K_0(t)
 \end{eqnarray}	
 for the velocity, kinetic energy $K(t)$, potential energy $ U(\textbf{r}^N(t))$ and energy $E(t)$ in MD. But 
  the total energy obtained by using Eq. (13) with $K(t)=K_0(t)$ for the kinetic energy 
  fluctuates with time although it remains constant, averaged over long time intervals.
 This is due to the fundamental quality of Newton's discrete dynamics, where the positions and
momenta appear asynchronous.

The energy invariance, $E_{\textrm{D}}$
in the discrete dynamics can, however,  be seen by considering the change in kinetic energy, $\delta K_ {\textrm{D}},$
potential energy,  $\delta U_ {\textrm{D}},$ and
the work done by the force 
in the time  interval $[t-\delta t/2, t+\delta t/2].$ The loss in  potential energy, $-\delta U_{\textrm{D}}$ is defined as
the work done by the forces at a  move of the positions \cite{Goldstein}. 
An expression for the work, $W_{\textrm{D}}$ done in the time interval by the discrete dynamics
from the position  $(\textbf{r}_i(t)+ (\textbf{r}_i(t-\delta t))/2$ at $t-\delta t/2$
to the position  $(\textbf{r}_i(t+\delta t)+ \textbf{r}_i(t))/2$ at $t+\delta t/2$ with the change in position
$\textbf(\textbf{r}_i(t+\delta t) -\textbf{r}_i(t-\delta t))/2$ is  \cite{Toxvaerd2014}
\begin{equation}
	-\delta U_{\textrm{D}}=W_{\textrm{D}}= \sum_i^N  \textbf{f}_i(t)(\textbf(\textbf{r}_i(t+\delta t) -\textbf{r}_i(t-\delta t))/2.
\end{equation}	
By rewriting Eq. (4) to
\begin{equation}
	\textbf{r}_i(t+ \delta t) -\textbf{r}_i(t-\delta t)= 2(\textbf{r}_i(t) -\textbf{r}_i(t-\delta t))+\frac{\delta t^2}{m_i} \textbf{f}_i(t),
\end{equation}
and inserting in Eq. (14) one obtains an expression for the total work in the time interval
\begin{equation}
	-\delta U_{\textrm{D}}=  W_{\textrm{D}}	=  \sum_i^N  [(\textbf(\textbf{r}_i(t) -\textbf{r}_i(t-\delta t)) \textbf{f}_i(t)+ \frac{\delta t^2}{2m_i}\textbf{f}_i^2].
\end{equation}

The change in kinetic energy in the time interval $[t-\delta t/2, t+\delta t/2]$ is
\begin{equation}
	\delta K_{\textrm{D}}= \sum_i^N \frac{1}{2}m_i [\frac{\textbf(\textbf{r}_i(t+\delta t) -\textbf{r}_i(t))^2}{\delta t^2}-
\frac{\textbf(\textbf{r}_i(t) -\textbf{r}_i(t-\delta t))^2}{\delta t^2}].
\end{equation}
(In time intervals without forces the dynamics follow Newton's first law.)
By rewriting  Eq. (4) to
\begin{equation}
	\textbf{r}_i(t+ \delta t) -\textbf{r}_i(t)= \textbf{r}_i(t) -\textbf{r}_i(t-\delta t)+\frac{\delta t^2}{m_i} \textbf{f}_i(t)
\end{equation}
   and inserting the squared expression for  $\textbf{r}_i(t+\delta t) -\textbf{r}_i(t)$ in  Eq. (17), the change in kinetic energy
   is
\begin{equation}
	\delta K_{\textrm{D}}= \sum_i^N [ (\textbf{r}_i(t)-\textbf{r}_i(t - \delta t))\textbf{f}_i(t) +\frac{\delta t^2}{2m_i} \textbf{f}_i(t)^2].
\end{equation}

The energy invariance in Newton's discrete dynamics is expressed by Eqn. (16),  and  Eq. (19) as \cite{Toxvaerd2014}
\begin{equation}
	\delta E_{\textrm{D}}=	\delta U_{\textrm{D}}+\delta K_{\textrm{D}}=0,
\end{equation}
where the difference in energies,  $ \sum_i^N \ \frac{\delta t^2}{2m_i} \textbf{f}_i(t)^2$, between the analytic and the discrete energies
 at time $t$ 
 only depends on the square of the forces and time increment.

In summary: Newton's discrete dynamics is time reversible and symplectic, and the discrete dynamics for a
conservative system has
the same invariances as the corresponding analytic dynamics. Newton's discrete dynamics is exact in the same sense as 
Classical Mechanics.

 \subsection{The connection between the discrete and  the analytic dynamics}

   \begin{figure}
	   \begin{center}	   
 	 \includegraphics[width=6.cm,angle=-90]{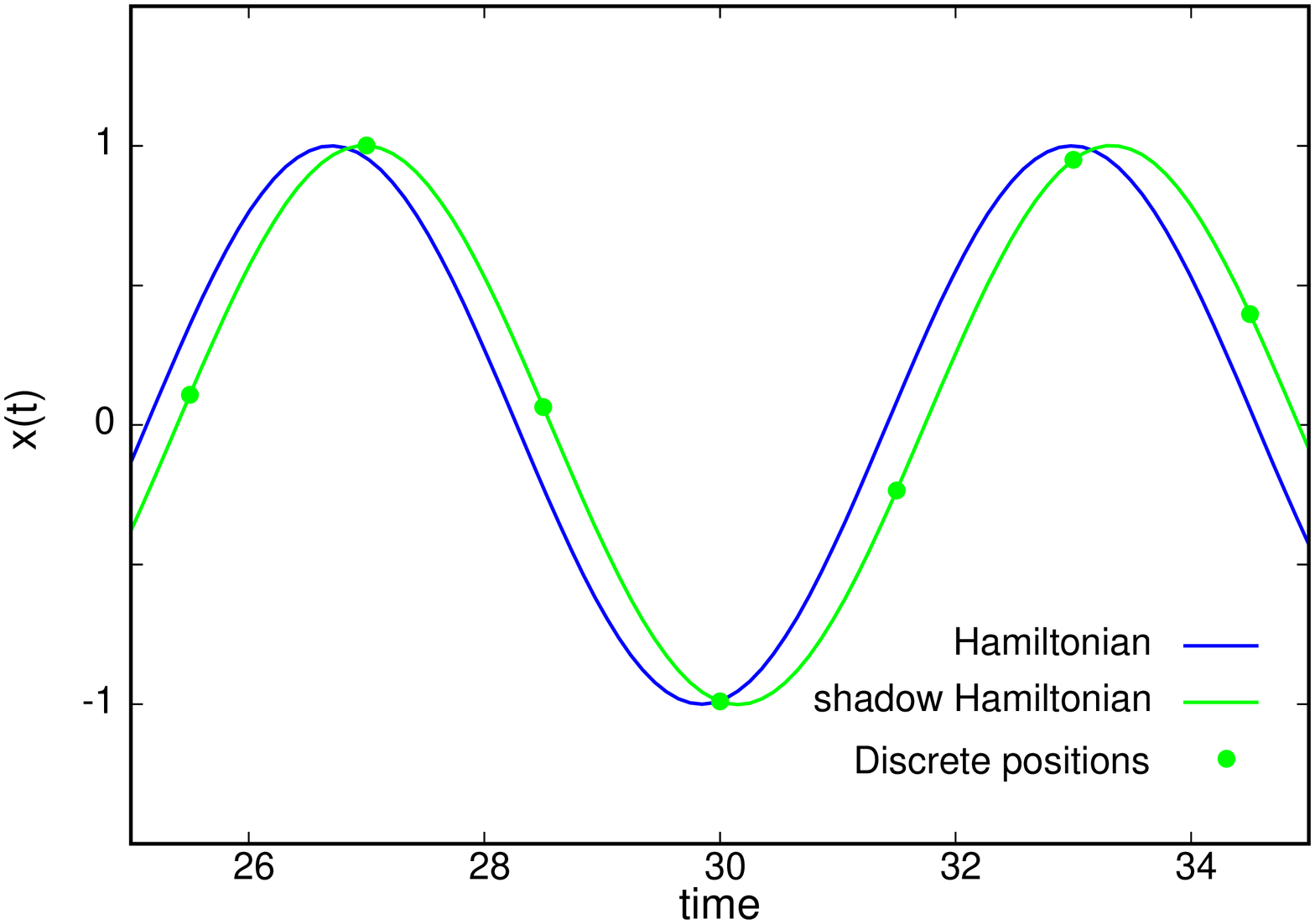}
	   \caption{ The dynamics for a one dimensioanal oscillator.
In blue is the analytic curve, Eq. (21), and the green curve is for the shadow Hamiltonian, Eq. (24).
The discrete positions with green dots are generated by Eq. (23) for a big time increment
 with $\approx$ four discrete positions per oscilllations.}
	   \end{center}		   
   \end{figure}
 The classical analytic dynamics, determined by the second order differential equation Eq.(6) can only be solved for a few systems, e.g. a  harmonic oscillator
 \begin{equation}
	 x(t)=A sin(\omega t).
 \end{equation}	 
 The corresponding difference equation for a discrete harmonic oscillator can, however, also be solved
 and the solution reveals the existence of a  \textit{shadow Hamiltonian}, $ \tilde{H} $ \cite{toxone}. 
 The discrete positions are located on a harmonic curve
 \begin{equation}
	 x(t)= \tilde{A}sin(\tilde{\omega} t).
 \end{equation}	

  The solution in \cite{toxone}  was obtained directly from the discrete
points and without any use of an  expansion of an analytic Hamiltonian $H(\textbf{r},\textbf{p})$  by noticing that if the  discrete dynamics
\begin{equation}
 x(t+\delta t)=2 x(t)- x(t-\delta t)- \omega^2 \delta t^2 x(t)=\alpha x(t) - x(t-\delta t),
\end{equation}
with $\alpha=2-\omega^2 \delta t^2$  is started with positions
 $x(0)=0$ and $x(\delta t)=A \sin(\omega \delta t)$, then
 the generated discrete points lie on a harmonic curve Eq. (22),
 with the frequency $\tilde{\omega}$  and amplitude $\tilde{A}$ given by
\begin{eqnarray}
	\tilde{\omega}=\cos^{-1}(1-\frac{(\omega \delta t)^2}{2})/\delta t \nonumber \\
	\tilde{A}=\frac{A \sin(\omega \delta t)}{\sin(\tilde{\omega} \delta t)},
\end{eqnarray}
 i.e. with the  harmonic shadow Hamiltonian,  $\tilde {H}(\tilde{\omega})$ with $ x(t)= \tilde{A}sin(\tilde{\omega} t)$
 and the energy
$\tilde {E}=(\tilde{A} \tilde{\omega})^2/2$. Eq. (24) sets, however, a limit to the discrete dynamics, given by $\delta t$.
 The discrete dynamics  are stable for
\begin{equation}
\mid 1-\frac{ \omega^2 \delta t_{max}^2}{2} \mid  \leq 1.
\end{equation}
or
\begin{equation}
 \delta t_{max} \leq \frac{2}{\omega}.
\end{equation}

Figure 3 shows the  positions of the one-dimensional harmonic oscillations in the time interval $t \in [25,35]$ after $\approx$ six oscillations from $x(0)=0$ at the
start at $t=0$.
The solution is for $ A=\omega=1$ and a
big time increment $\delta t=1.5< 2/\omega=2$, for which
$ \tilde{A}= 1.00119889$ and $\tilde{\omega}=0.98983513$, and the energies are $E=0.5$ and $\tilde{E}=0.49106$, respectively.
The discrete positions (green dots) are obtained by  the discrete dynamics Eq. (23), and the green curve is $x(t)=\tilde{A} sin(\tilde{\omega})$
for the shadow Hamiltonian $\tilde{H}(\tilde{\omega})$. The blue curve is the harmonic solution Eq. (21).
The discrete dynamics only generates $\approx$ 4 positions per oscillation.

The existence of a shadow Hamiltonian is a general
property of discrete symplectic dynamics.
Mathematical investigations have established \cite{Griffiths,Sanz-Serna,Hairer,Reich}
 that, if a discrete algorithm is symplectic, then there 
exists  a shadow Hamiltonian,  $ \tilde H(\textbf{r},\textbf{p})$ 
 for sufficient small  $\delta t$ such that the discrete positions
$\textbf{r}(n\delta t)$, for an object, generated by the symplectic algorithm,
 lie on the analytic trajectory for the object obtained by  $\tilde H(\textbf{r}(t),\textbf{p}(t))$.

Newton's discrete algorithm, Eq. (3), is symplectic and the first
  non-trivial term in the asymptotic expansion was obtained for the LJ system in \cite{toxone}.
 The  term was obtained from consecutive sets of positions
using the expression obtained from the expansion of $H$ for a one-dimensional harmonic oscillator.
 Reference \cite{Gans} gives the general expression for the first term in the expansion using the method of modified equations
 derived in Refs. \cite{Griffiths,Sanz-Serna,Hairer,Reich}. The Hessian, $\partial^2  U(\textbf{r}^N)/\partial \textbf{r}^2$,
  of the potential energy function $U(\textbf{r}^N)$ is denoted by  $\textbf{J}$, the velocities
 of the $N$ particles by
 $\textbf{v}^N\equiv(\textbf{v}_1, ..., \textbf{v}_N)$, and the force 
 at positions  $\textbf{r}^N$ by
 $\textbf{F}^N(\textbf{r}^N)\equiv(\textbf{f}_1(\textbf{r}^N), ..., \textbf{f}_N(\textbf{r}^N))$.
 Using Eq. (11) for $\textbf{v}^N$, the first term in the asymptotic expansion at the $n$'th time step 
 can be expressed  as \cite{toxtwo}
\begin{eqnarray}
	\tilde{E}(t_n) \simeq E_0(t) +
	\frac{\delta t^2}{12} (\textbf{v}_n^N)^T
 \textbf{J(r}_n^N) \textbf{v}_n^N - \frac{\delta t^2}{24m}\textbf{F}_n^N(\textbf{r}_n^N)^2 \nonumber \\
	= E_0(t)+E_1(t)
\end{eqnarray}
for the  the energy, $E_0+E_1$, at time $t=n \delta t$.
The zero-order term is the discrete energy $E_0$ (Eq. (13)) used in MD. The detailed expression for the first order term $E_1$ in the
asymptotic expansion for a system with pair interactions is given in \cite{toxtwo}.
 
   \begin{figure}
	   \begin{center}	   
 	 \includegraphics[width=6.5cm,angle=-90]{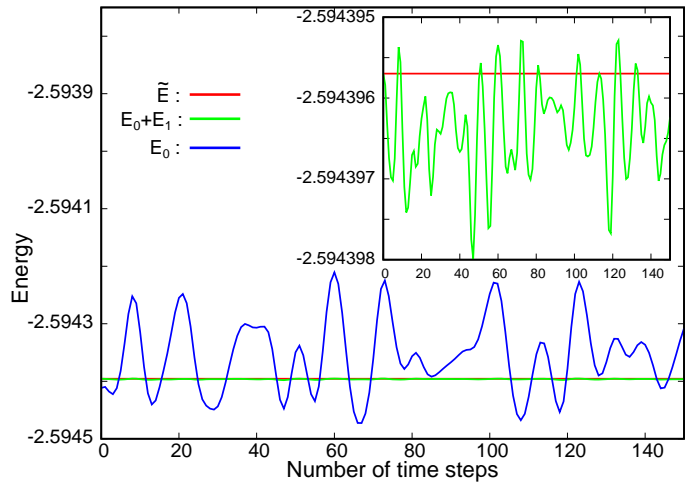}
	   \caption{ Energies at 150 time steps with the discrete dynamics for a system of Lennard-Jones particles.
	The traditional energy $E_0(t)$ used in
	    MD simulations is shown in blue. The first order expression for the shadow energy $E_0(t)+E_1(t)$,
	    Eq. (27),  is shown in green, and the constant shadow energy
	    energy $\tilde{E}(t_0)+\Sigma_n \delta E(n \delta t)$ , Eq. (20), is shown with red. The energies  $E_0(t)+E_1(t)$ and  $\tilde{E}(t_0)+\delta E(t)$ are enlarged in the inset.}
	   \end{center}		   
  \end{figure}

 The existence of a shadow Hamiltonian for  classical dynamics  can be demonstrated by MD for
 a system of  particles with Lennard-Jones  (LJ) pair interactions
 \begin{equation}
	 u(r_{ij})= 4\epsilon[(r_{ij}/\sigma)^{-12}-	(r_{ij}/\sigma)^{-6}].	 
 \end{equation}

 The energy conservation in discrete dynamics for a system of $N$ particles with LJ interactions is shown in Figure 4.
 The energies are for  a system of $N=2048$
 particles 
 in a liquid state with
  the number density $\rho=0.80$ and temperature $T=1.00$
  (for units of length, time, and energy in MD see \cite{MDunits}). The energies in the figure are for 150 time steps with $\delta t=$ 0.005 and
  shown from  a  configuration (at $t=t_0$) where $E_0(t_0) \approx E_0(t_0)+E_1(t_0)$.
 The energy estimate $E_0(t)$ used in
	   almost all MD simulations is shown in blue. The first order expression Eq. (27) \cite{toxone,toxtwo}) is shown in green,
	   and the constant  energy    $\tilde{E}(t)=\tilde{E}(t_0)+\Sigma_n \delta E(n\delta t) $ of the shadow Hamiltonian for the discrete dynamics
	   \cite{Toxvaerd2014}  is with red. (The shadow energy $\tilde{E}(t_0)$   is not known, and $\tilde{E}(t)$ in the figure is 
	    $E_0(t_0)+E_1(t_0)+\delta E(n \delta t)$ with $\delta E(n \delta t)$ given by Eqn. (14) and (17)).
	   The first order correction $E_1(t)$ of the energy decreases the fluctuations in the energy
	   with a factor of hundred \cite{toxone,toxtwo} and the difference $\delta (\tilde{E}(t)-(E_0(t)+E_1(t)))$ from the energies of
	   the higher-order term in the expansion can not be seen in the figure but is visible in the inset.

\begin{figure}
\begin{center}
\includegraphics[width=6.cm,angle=-90]{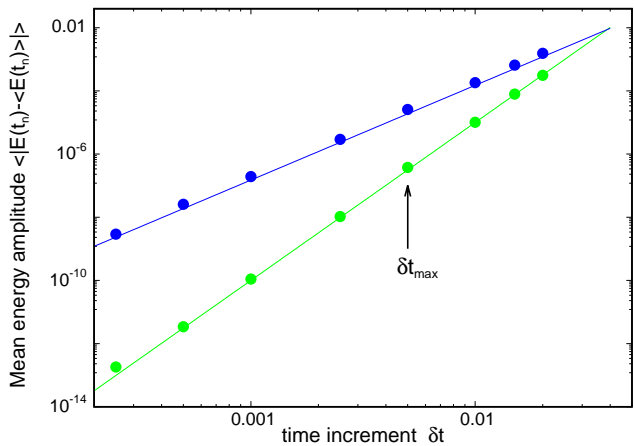}
	\caption{The mean  amplitudes, $<|E(t_n)-<E(t_n)>|>$, of the  energy in a LJ system at  $\rho=0.80$ and $T=1.00$
  as a function of the time-increment $\delta t$ in a $\it{log}-\it{log}$ plot.
	The green dots are for the first order estimate, Eq. (27), of  the shadow energy  $E_0+E_1$ and 
	blue dots are for $E_0$. The 
	 lines through the green and blue dots have slopes of five and three, respectively. $\delta t_{\textrm{max}}=0.005$ is used in the first discrete 
	 Newtonian MD simulation in 1967 by L. Verlet \cite{Verlet1967}.}
\end{center}
\end{figure}
 Simulations for different state points and time increments indicate,
	   that the energy at the  asymptotic expansion rapidly converges  towards $\tilde{E}(\rho,T)$ for time increments
	   used in MD \cite{toxtwo}. In Figure 5  the means of the energy  amplitudes $<|E(t_n)-<E(t_n)>|>$ in the LJ system,
	   (the amplitudes for $\delta t=0.005$ are shown in Figure 4)
	   are plotted as a function
	   of the time increment $\delta t$. The green dots are the amplitudes for $E(t_n)= E_0(t_n)+ E_1(t_n)$ and the blue dots are for
 $E(t_n)= E_0(t_n)$.   Systems of liquids of LJ  particles  or mass units in molecules for e.g. carbon atoms in organic molecules  have  ''vibration times"
 of the order $\approx 10^{-12}$ sec, and the maximum time increment used in these MD simulations are
 $\delta t_{\textrm{max}} \approx 0.005 \approx 10^{-14}$ sec, used by L. Verlet in 1967 in the first discrete Newtonian MD \cite{Verlet1967}.
 It corresponds to of the order a hundred discrete positions for a typical
 ''vibration"; but most MD simulations are, however, with smaller time increments.  The data in Figure 4 and Figure 5 indicate that
 the asymptotic expansion is rapidly converging for these values of the time increment and that
 the energies are well determined by $E_0(t_n)+ E_1(t_n)$ as well as by  $E_0(t_n)$.

In summary: There exists most likely a shadow Hamiltonian for discrete molecular dynamics, and
the asymptotic expansion is rapidly converging for the 
time increments used in discrete molecular dynamics. The existence of a shadow Hamiltonian implies
that there is no qualitative difference between the analytic Classical Mechanics and the discrete
 Newtonian dynamics. The two kinds of dynamics are, however, mathematically different (see next subsection).

\subsection{The difference between  Classical Mechanics and  discrete Newtonian dynamics }

The dynamics obtained by Classical Mechanics or discrete Newtonian dynamics are qualitatively equal,
despite the fact that they mathematically are fundamentally different. The difference between the two dynamics can perhaps
best be seen by rewriting  $\textit{Proposition I}$

\begin{equation}
	\textbf{r}(t+\delta t)-	\textbf{r}(t)=\textbf{r}(t)-	\textbf{r}(t-\delta t)  +\delta t^2 \textbf{f}(\textbf{r}(t))/m,	
					 \end{equation}
and					 
\begin{equation}
	\textbf{r}(t+\delta t)=\textbf{r}(t)+\delta t(\textbf{r}(t+\delta t)-	\textbf{r}(t))
					 \end{equation}
 where the new position $\textbf{r}(t+\delta t)$ is obtained  from the old position $  \textbf{r}(t)$ by   
  that  the difference with the previous positions 
   is adjusted by a
  force quantum  $\delta t^2 \textbf{f}(\textbf{r}(t))/m$. Or as Newton expressed it:
\textit{..suppose that a centripetal force acts at once
with a great impulse}. The time and space are discretized    in $\textit{Proposition I}$ 
and connected by a quant of the square of the time quant and the ratio between the force and the mass.

\textit{ It is only the positions, the time, and the force that appear in the dynamics, Eqn. (29) and (30).
The momenta are not dynamic variables in
discrete dynamics.}

Although Newton starts $Principia$ with his first law for momenta and refers to the law in the formulation of $Proposition$ $I$,
the momenta do not enter into the dynamics, but they play a role as a ''bookkeeping" registration during the time evolution.
In Newton's analytic dynamics, the momenta are, however, also dynamic variables, and due to this fact, the
analytic dynamics can be reformulated to  Lagrangian mechanics \cite{Lagrange1788}.

Classical Mechanics is the analytic limit dynamics  of nonrelativistic quantum mechanics \cite{Wigner}, and the question is: Could 
the discrete Newtonian dynamics be a corresponding limit dynamics of discrete nonrelativistic quantum mechanics?
Noble Laureate T. D. Lee wrote in 1983 a paper  entitled, "Can Time Be a Discrete Dynamical Variable?"\cite{Lee1}.
The article led to a series of publications by Lee and collaborators on the formulation of fundamental
dynamics in terms of difference equations, but with exact invariance under
continuous groups of translational and rotational transformations.
Quoting Lee \cite{Lee2}, he "wish to explore an alternative point of view: that physics
should be formulated in terms of difference equations
and that these difference equations could exhibit all the desirable symmetry properties and conservation laws".
 Lee's analysis covers not only classical mechanics \cite{Lee1}, but also non-relativistic quantum mechanics
 and relativistic quantum field theory  \cite{Lee3}, and  Gauge theory and Lattice Gravity \cite{Lee2}.
Lee's discrete dynamics is obtained by treating only the positions but not the momenta as  discrete dynamical variables,   as
is the case in discrete Newtonian dynamics.

The discrete nonrelativistic quantum mechanics is obtained by Lee using Feynman's path
 integration formalism, but for  discrete positions and a corresponding 
discrete action,
\begin{equation}
	\mathcal{A}_D = \sum_{n=1}^{\mathcal{N} +1}(t_n-t_{n-1}) \left [ \frac{1}{2}\frac{(\textbf{r}^N_{n}-\textbf{r}^N_{n-1})^2}{(t_n-t_{n-1})^2}+
\overline{V(n)} \right ],
\end{equation} 
where $\textbf{r}^N_{\mathcal{N}+1}$ is the end-positions  at time $t_{\mathcal{N}+1}$ and the  minimum  of $ \mathcal{A}_D$ determines the classical path.
According to Lee, the  action is a sum over products of time increments and energies with "kinetic energies" 
 and $ \overline{ V(n)}$, for the 
  average of "potential energy" in the
time intervals $[t_{n-1},t_n]$. In discrete Newtonian dynamics, it is the action sum over products of time intervals
and with the classical limit  energies given by
the shadow Hamiltonians in the time intervals 
\begin{equation}
	\tilde{E}(t)= 	 \left [ \frac{1}{2}\frac{(\textbf{r}^N_{n}-\textbf{r}^N_{n-1})^2}{(t_n-t_{n-1})^2}+
\overline{V(n)} \right ].
\end{equation}

In Lee's formulation of discrete mechanics, "there
is a $\textit{fundamental length l}$ or time $t_l$ (in natural units). Given any time interval
 $T=t_{\textrm{f}}-t_0$, the total number $\mathcal{N}$ of discrete points that define the
trajectory is given by the integer nearest $T/l$."
 The classical discrete trajectory is the
 classical limit path for discrete quantum mechanics
 with $\delta t=t_l$, as the classical analytic trajectory is for traditional quantum mechanics.
There is, however, one important difference between analytic and discrete dynamics.
 The momenta   $\textit{for all the paths}$ in the discrete quantum dynamics aŕe not dynamic variables. They are obtained by a difference between
discrete sets of positions and they are all $\textit{asynchronous}$ with the positions. So
the Heisenberg uncertainty principle is a trivial consequence of  Lee's  discrete quantum dynamics.
  The fundamental length and time  in  quantum electrodynamics (QED) are
the  Planck length $l_{\textrm{P}} \approx 1.6 \times 10^{-35}$m and
 Planck time $t_{\textrm{P}} \approx 5.4 \times 10^{-44}$ s \cite{Garay},
and they are immensely smaller than the length unit 
 and time increments used in MD to generate the classical discrete dynamics.
 But the analogy implies that the discrete  Newtonian dynamics  is the classical limit  of the
Lee's discrete non-relativistic quantum mechanics.

 In Newton's discrete dynamics, the contributions from the forces are also discretized. T. Regge and R. M. Williams have   analyzed 
the dynamical implications of a discretized gravitational force \cite{Regge2000}.

In summary: Newton's discrete and analytic dynamics are mathematically different. The fact that the positions and momenta are asynchronous
implies, that it could be that discrete dynamics in principle is the correct classical limit dynamics of a discrete quantum dynamics.
But the  energy difference, Eq. (20), between the two kinds of classical dynamics are, however, proportional to $ t_{\textrm{P}}^2$ and incredibly small, and since there is no qualitative difference
between the two classical dynamics, there is a lack of justification for preferring the discrete quantum mechanics.

\section{Discrete Molecular Dynamics}
Almost all MD simulations are performed with Newton's discrete algorithm, Eq. (3). The MD algorithm  is, however, hardly mentioned
 in the simulations.
 The simulations are stable at the time propagation, and the focus in the articles is on the physics of the MD systems rather that on the computational method.
 And with good reason because the algorithm offers an exact time propagation of the MD models provided that the time step is not too big so that
 the asymptotic expansion no longer converges. The data of the simulated models depends primarily on the quality of the expressions for the forces
 and the molecular mechanics force fields used in the simulations,  and not on the
 discrete dynamics.

 There are, however, some factors that can affect the exactness of the time propagation.
 The positions are in the best cases  given in ''double-precision" with the order 
 $\approx 10^{-16} $ relative accuracy, and the round-off errors in long  simulations generate cumulative errors in the numerical integration.
 The accumulations will, however, normally not affect the results obtained by the simulation \cite{Toxvaerd1991}, and the round-off errors can optionally be avoided by
 performing the MD simulation with integer arithmetic \cite{Levesque1993}.

 Most MD simulations contain approximations of the forces and the size of the systems with the result, that the
 simulations no longer are exact, but have to be adjusted e.g. for energy conservation.  The range of the interactions
 is often set to zero at a sufficiently large distance, and this approximation causes a small drift in the energy.
 The interactions are usually restricted at these large distances by a ''cut and shift" of the potentials, but since it is the forces that enter into the algorithm it
 is much better to cut and shift the range of the forces \cite{Toxvaerd2011}. A cut in the potentials introduces new ''Delta function" forces in the systems and they perform
 a work on the system according to Eq. (16).
 This will affect the energy conservation and show up as a small drift in the temperatures of the systems unless one uses a thermostat,
 which one, however, usually do.

 The MD computational method is described in \cite{Tildesley,Frenkel}.  This section is ended by showing
 some examples where the exactness of the dynamics is used to obtain the exact dynamics of complex systems. The differential equation(s) for the   analytic dynamics in Classical Mechanics
 can only be
 solved for a few simple systems, but the discrete Newtonian dynamics with the algorithm, Eq. (3), offers a general method for obtaining exact discrete
 solutions, e.g. regular solutions of complex systems of celestial objects. Three examples are given here: $4.1$  The emergence and evolution of planetary systems.
$4.2$ The emergence and evolution of planetary systems with inverse forces.  $4.3$ The emergence and evolution of galaxies in the expanding Universe.

\subsection{The emergence and evolution of a planetary system}

According to Newton's shell theorem \cite{Newtonshell} the force, $\textbf{F}_i$, 
on  a spherically symmetrical object $i$ with mass $m_i$ is a sum over the forces, $ \textbf{f}(r_{ij})$, caused by the other 
spherically symmetrical objects $j$ with mass $m_j$, and it
is  solely  given by their center of mass distance $r_{ij}$ to $i$
 \begin{equation}
	 \textbf{F}_i(r_{ij})= \sum_{j \neq i}^N \textbf{f}(r_{ij})=- \sum_{j \neq i}^N \frac{G m_i m_j}{r_{ij}^2}\hat{\textbf{r}}_{ij}.
 \end{equation}

Newton's discrete algorithm can be extended to include a ''perfect" fusion of mass objects \cite{Toxvaerd2022}.
Let all the spherically symmetrical objects
 have the same (reduced)  number density $\rho= (\pi/6)^{-1} $ by which
 the diameter $\sigma_i$ of the spherical object $i$ is 

 \begin{equation}
	 		 \sigma_i= m_i^{1/3}
 \end{equation}
  and   the collision diameter 
\begin{equation}
	\sigma_{ij}=	\frac{\sigma_{i}+\sigma_{j}}{2}.
\end{equation}	
 If  the distance $r_{ij}(t)$ at time $t$ between two objects is less than $\sigma_{ij}$ 
the two objects merge into one spherical symmetrical object with mass

\begin{equation}
m_{\alpha}= m_i + m_j,
\end{equation}	 
and diameter
\begin{equation}
 \sigma_{\alpha}= (m_{\alpha})^{1/3},
\end{equation}
and with the new object $\alpha$  at the position
\begin{equation}
	\textbf{r}_{\alpha}= \frac{m_i}{m_{\alpha}}\textbf{r}_i+\frac{m_j}{m_{\alpha}}\textbf{r}_j,
\end{equation}	
at the center of mass of the two objects before the fusion.
(The   object $\alpha$ at the center of mass of the two merged objects $i$ and $j$ might occasionally be near another object $k$
by which more objects merge, but after the same laws.)
Let the center of mass of the system of  the $N$ objects be at the origin, i.e.
\begin{equation}
			\Sigma_k m_k \textbf{r}_k(t)=\textbf{0}.
\end{equation}

The momenta  of the objects in the discrete dynamics just before the fusion are $\textbf{p}^N(t-\delta t/2)$ and the
total momentum of the system is conserved  at the fusion if
\begin{equation}
\textbf{v}_{\alpha}(t-\delta t/2)= \frac{m_i}{m_{\alpha}}\textbf{v}_i(t-\delta t/2)+ \frac{m_j}{m_{\alpha}}\textbf{v}_j(t-\delta t/2),
\end{equation}
which determines the  velocity $\textbf{v}_{\alpha}(t-\delta t/2)$ of the merged object.

The invariances in the classical
Newtonian dynamics are for a conservative system with Newton's third law, i.e with
\begin{equation}
	\textbf{f}_{kl}(t)=-\textbf{f}_{lk}(t)
\end{equation}	
for the forces between two objects $k$ and $l$, and with no external forces.
An  object $k$'s forces with $i$ and $j$ before the fusion 
 are   $\textbf{f}_{ik}(t)$ and $\textbf{f}_{jk}(t)$,
and these forces
must be replaced by calculating the force $\textbf{f}_{\alpha k}(\textbf{r}_{\alpha k}(t))$.
The total force after the fusion is zero 
due to Newton's third law  for a conservative system with  the forces $\textbf{f}_{\alpha k}=-\textbf{f}_{k \alpha}$ between pairs of objects,
 and the total momentum
\begin{eqnarray}
	\Sigma_k \textbf{p}_k(t_n+\delta t/2)= \Sigma_k \textbf{p}_k(t_n-\delta t/2)+ \delta t\Sigma_k \textbf{f}_k(t_{n}) \nonumber \\
	= \Sigma_k \textbf{p}_k(t_n-\delta t/2),
\end{eqnarray}	
and the  position of the center of mass, Eq. (39), is conserved for the discrete dynamics with fusion.

 The determination of the position, $\textbf{r}_\alpha(t)$, and 
 the velocity, $\textbf{v}_\alpha(t-\delta t/2)$, of the new object from the requirement of  
 conserved center of mass and conserved momentum determines the discrete dynamics of the $N-1$ objects.

 The total angular momentum is also not affected by the fusion.
 The angular momentum of the system of  spherically symmetrical objects consist of two terms
 \begin{equation} 
	 \textbf{L}(t)= \textbf{L}_{G}(t)+ \textbf{L}_{S}(t)
 \end{equation}
 where $ \textbf{L}_{G}(t)$ is the angular momentum of the  objects in their orbits, due to the dynamics obtained from the gravitational ($G$) forces between their
 centers of mass, and $\textbf{L}_{S}(t)$ is the angular momentum  due to the  spin ($S$) of the objects. 
 Without fusion  $\textbf{L}_{G}(t)$    is  conserved for Newtons discrete
dynamics, Eq. (10).    $\textbf{L}_{S}(t)$ is, however, also  conserved  according  to the shell theorem \cite{Newtonshell} , where Newton
proves that  no net gravitational force is exerted by
a shell on any object inside, regardless of the object's location within the uniform shell, by which the
spin of the object is not affected by any force and is therefore constant. 
But at a fusion $ \textbf{L}_{G}$ changes  by
\begin{eqnarray}
	\delta \textbf{L}_{G}(t)= \nonumber \\
	\textbf{r}_{\alpha}(t) \times m_{\alpha}\textbf{v}_{\alpha}(t-\delta t/2)- \nonumber \\
 \textbf{r}_i(t) \times m_i\textbf{v}_i(t-\delta t/2)- \textbf{r}_j(t) \times m_j\textbf{v}_j(t-\delta t/2).
\end{eqnarray}	 
and $ \textbf{L}_{S}$ changes  by
\begin{eqnarray}
\delta \textbf{L}_{S}(t)= \nonumber \\
(\textbf{r}_i(t)- \textbf{r}_{\alpha}(t))\times m_i\textbf{v}_i(t-\delta t/2)+ \nonumber \\
(\textbf{r}_j(t)- \textbf{r}_{\alpha}(t)) \times m_j\textbf{v}_j(t-\delta t/2) \nonumber \\ 
= \textbf{r}_i(t) \times m_i\textbf{v}_i(t-\delta t/2)+ \textbf{r}_j(t) \times m_j\textbf{v}_j(t-\delta t/2)  \nonumber \\
 - \textbf{r}_{\alpha}(t) \times m_{\alpha}\textbf{v}_{\alpha}(t-\delta t/2) \nonumber \\
=	-\delta \textbf{L}_{G}(t).	
\end{eqnarray}	 

So  without fusion, the angular momenta $ \textbf{L}_{S}(t)$ and $ \textbf{L}_{G}(t)$ with Newton's discrete dynamics  are
conserved separately, and at a fusion, the  total angular momentum is still  conserved but with an exchange of angular momentum with
$\delta \textbf{L}_{S}(t)= -\delta \textbf{L}_{G}(t)$.
\begin{figure}
\begin{center}
\includegraphics[width=6.cm,angle=-90]{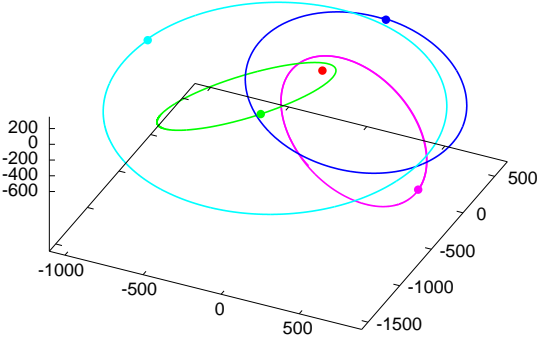}
	\caption{ The regular orbits of the four innermost   planets in a planetary system. The planetary system contains 21 planets in regular orbits \cite{Toxvaerd2022}.
	The orbit times (MD time unit) $t_{\textrm{orbit}}$ and eccentricity $\epsilon$ are:  light blue: $t_{\textrm{orbit}}=818, \epsilon=0.835$; green: $t_{\textrm{orbit}}=300, \epsilon=0.327$;
	blue	$t_{\textrm{orbit}}=463, \epsilon=0.586$; magenta $t_{\textrm{orbit}}=313, \epsilon=0.768$. The green planet has circulated more than four thousand times around
	the Sun (red). }
\end{center}
\end{figure}

The exact classical discrete dynamics with a fusion of colliding objects can be used to explore the self-assembly
at the emergence of planetary systems and to investigate the stability of a planetary system. The emergence and
stability of planetary systems were investigated in \cite{Toxvaerd2022}, and Figure 6 shows the orbits 
of the four innermost planets at the end of the simulation
in a planetary system with 21 planets in regular orbits around a heavy gravity center. The emergence of the
simple planetary system was obtained from thousand objects
relatively close to each other and with a Maxwell Boltzmann distribution of the velocities
and with the dynamics given by the Eqn. (5),(34)-(38) and
(40).
For the setup and details see  \cite{Toxvaerd2022}.

\subsection{The emergence and evolution of planetary systems with inverse forces }

 Newton was aware  that the extension of an object 
  can affect the gravitational force between two objects, and in $\textit{Theorem XXXI} $ in $Principia$ \cite{Newtonshell}  
he solved this problem for the gravitational inverse square forces (ISF), Eq. (33), between spherically symmetrical objects.
 Newton's theorem is, however, only valid for ISF. But for inverse gravitational forces (IF) and inverse cubic gravitational forces (ICF)
 one can expand the contributions in powers of the ratio $\sigma/r$ between the diameters of the objects and their mutual distance. The
 result is \cite{Toxvaerd2022a}.
 
For the IF gravitational forces:
\begin{equation}
	\mathbf{f}_{ij}(r_{ij}) \simeq -\frac{G_1 m_im_j}{r_{ij}}(1 -\frac{\sigma_i^2+\sigma_j^2}{5r_{ij}^2})\hat{\mathbf{r}}_{ij}+ \mathcal{O}(r_{ij}^{-4}) 
\end{equation}
For  ISF one obtains the usual expression for the gravitational  forces  which do not depend on the extensions of the two spherically symmetrical objects
and is given by Eq. (33).

\begin{figure}
\begin{center}
\includegraphics[width=8cm,angle=-90]{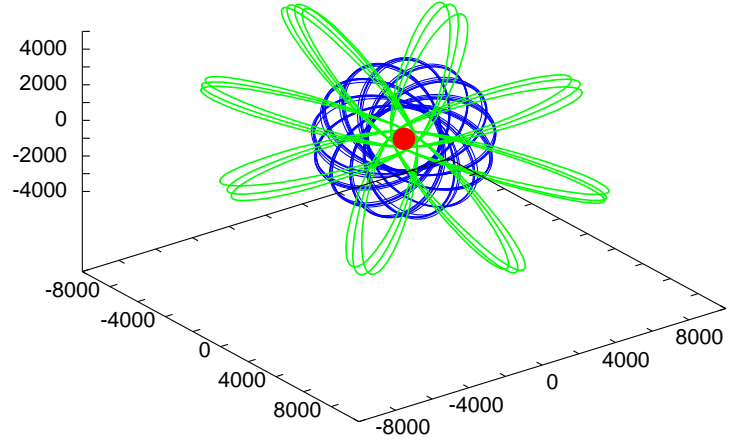}
	\caption{ Regular orbits for two of the 38 planets in a planetary system with inverse force attraction \cite{Toxvaerd2022a}.
	The orbits with green are for twenty-five ''revolving orbits" of one of the  planets in the system, and the orbits in blue are correspondingly sixty-three
	revolving orbits of another planet. With red is the (enlarged) center of mass of the planetary system.}
\end{center}
\end{figure}

For   the ICF  gravitational forces:
\begin{equation}
	\mathbf{f}_{ij}(r_{ij}) = -\frac{G_3 m_im_j}{r_{ij}^3}(1 +\frac{2\sigma_i^2+2\sigma_j^2}{5r_{ij}^2})\hat{\mathbf{r}}_{ij}+ \mathcal{O}(r_{ij}^{-6}). 
\end{equation}

Planetary systems with IF attractions were created in the same way as for the usual ISF attractions \cite{Toxvaerd2022a},
and the systems were even easier to create.
There is, however, a remarkable difference between the regular orbit in an ordinary planetary system and the regular orbits of objects in a planetary system
with IF forces. Whereas the orbits of the planets in e.g. our Solar system are regular with elliptical orbits, the orbits in an IF planetary system is what Newton probably
would have called ''revolving orbits". An example is shown in Figure 7. The figure shows the orbits
at the end of the simulation ($t=2.5 \times 10^6$) of two of the planets in a planetary system with
38 planets. All the planets have performed many orbits around the common center of gravity (the ''Sun" enlarged and with red in the figure), but none of the orbits were elliptical.
They showed different kinds of revolving orbits where the ''orientation" of the regular orbit was changed when the planet passed the ''Perihelion". The orientation
of the ''principal axis" was then changed with $\approx$  fractions of $2 \pi$. For details see \cite{Toxvaerd2022a}.

The Moon exhibits apsidal precession, which is called Saroscyclus and it has been known since ancient times. 
  Newton shows in Proposition 43-45 in $Principia$,  that the added force on a single object from a fixed mass
  center which can cause its apsidal precession  must be a central force
  between the planet and a  mass point fixed in space (the Sun). In Proposition 44 he  shows
  that  an inverse-cube force (ICF) might cause the revolving orbits, and  in Proposition 45
  Newton extended his theorem to arbitrary central forces by assuming
  that the particle moved in a nearly circular orbit \cite{Chandrasekbar}.
  The Moon's apsidal precession is explained by flattering by the rotating Earth with tide waves, which
  causes an ICF on the Moon. For Newton's analysis of the Moon's apsidal precession see \cite{Aoki1992}.

  It was, however, not possible to obtain stable regular orbits for systems with pure ICF, but systems with  the  gravitational ISF attraction, perturbed
  by additional ICF showed revolving orbits in qualitative accordance with the dynamics of the Moon.

  The conclusion from the investigation in \cite{Toxvaerd2022a} is that the gravitational ISF is the limit of attraction, $-G_n m_i m_j/r^n$ with respect to the power $n$.
  A system of $N$ objects can only have regular orbits for $n \leq 2$.

\subsection{The emergence and evolution of galaxies in the expanding Universe}

The dynamics of galaxies in an expanding universe are often obtained for gravitational  and dark matter in an Einstein-de Sitter universe  \cite{Vogelsberger2019},
or alternatively, by modifying the weak accelerations from gravitation long-range attractions (MOND) \cite{Milgrom1983}.
But the time evolution of galaxies
can also be determined for galaxies with pure gravitational forces by discrete Newtonian dynamics \cite{Toxvaerd2022b}.
The time-reversible algorithm  for the formation and aging of gravitational systems by self-assembly of baryonic objects in section 4.1
is extended to include the Hubble expansion of the space \cite{Hubble1929}. The algorithm is stable
for billions of time steps without any adjustments. The algorithm was used to simulate simple models of the Milky Way with the Hubble
expansion of the universe, and the galaxies were simulated for times that correspond to more than 25 Gyr.

The expansion of the space during the discrete Newtonian time propagation of a galaxy was obtained by comparing it with the Hubble expansion of the Universe.
A galaxy, $l$, far away in the universe moves away from the Earth at a speed proportional to its ''proper  distance", $H r_{kl}(t)$,
to the Earth, $k$, measured at the  ''cosmological time" $t$, and this behaviour 
is explained by an expanding universe.
The Hubble constant $H$ is the expansion coefficient in Hubble's law  \cite{Hubble1929,Lan2022}
\begin{equation}
  	\textbf{v}^{H}(\textbf{r}_{kl}(t))  =H \textbf{r}_{kl}(t)
\end{equation}
for the velocity, $	\textbf{v}^{H}(\textbf{r}_{kl}(t)) $ of a galaxy $k$ a  distance $ \textbf{r}_{kl}(t)$ from the Earth $l$.
The Hubble expansion can be obtained by an intensive expansion of the space 
 independently of the baryonic matter in the universe
\begin{equation}
	 	\textbf{v}^{H}(\textbf{r}(t))=H \textbf{r}(t),
 \end{equation}
 by which the distance between pairs of positions $\textbf{r}_{k}(t),\textbf{r}_{l}(t),$ or $\textbf{r}_k(t),\textbf{r}_k(t+\delta t)$ increases with
 the Hubble velocity.  Eq. (49) fulfills   the \textit{Cosmological principle} and the 
 \textit{Copernican principle} for expansion of the universe \cite{Book}. The Cosmological principle, which was first formulated by Newton \cite{Newton1687}  demands,
 that no place in the universe is preferred,
 and the Copernican principle 
 demands, that no direction in the universe is preferred.   Eq. (49) will accelerate the expansion of the universe, and 
 observations of galaxies indicate, that the distances between the galaxies accelerate (dark energy)\cite{Riess1998,Perlmutter1999}.

The Hubble constant $H$ is  quoted in $\textrm{km s}^{-1}\textrm{Mpc}^{-1}$, for the velocity in $\textrm{km s}^{-1}$ 
of a galaxy 1 megaparsec ($3.09 \times 10^{19}$ km) away.
Its value is  \cite{Soltis2021}
\begin{equation}
	H=72.1 \pm  2.0 \  \textrm{km} \ \textrm{s}^{-1} \textrm{Mpc}^{-1}.
\end{equation}
The Hubble velocity has no effect on the orbits of the planets and the stability of the Solar system, but it affects the stability of galaxies.

The Hubble expansion is included in the discrete Newtonian dynamics \cite{Toxvaerd2022b}.
Newton's discrete dynamics
changes the position of an object $k$.
If the space expands monotonously over time with
the Hubble velocity $\textbf{v}^H$, then the expansion also changes the distance between two positions.  The new position  $\textbf{r}_k(t+\delta t)$ is the
sum of the change due to the gravitational force on $k$ and the contribution from the Hubble expansion.
The mean location of an object   changes  from
$\textbf{r}_k(t-\delta t/2)=(\textbf{r}_k(t-\delta t)+\textbf{r}_k(t))/2$ at $t \in[t-\delta t, t]$ to
$\textbf{r}_k(t+\delta t/2)=(\textbf{r}_k(t)+\textbf{r}_k(t+\delta t))/2$ at  $t \in [t, t+ \delta t]$  .
The Hubble expansion changes the distance between the two positions by
the Hubble velocity
\begin{eqnarray}
\textbf{v}^H= H \delta \textbf{r}_k = H(\textbf{r}_k(t+\delta t/2)-\textbf{r}_k(t-\delta t/2)) \nonumber \\
=\delta tH \frac{\textbf{r}_k(t+\delta t)-\textbf{r}_k(t)}{2\delta t}+
\delta tH \frac{\textbf{r}_k(t)-\textbf{r}_k(t-\delta t)}{2\delta t} \nonumber  \\
=\frac{\delta t H}{2} \textbf{v}_k(t+\delta t/2) + \frac{\delta t H}{2} \textbf{v}_k(t-\delta t/2).
\end{eqnarray}
By including the Hubble velocity, Eq. (51),  in the
Newtons algorithm, Eq. (5), and after a re-arrangement,
one obtains the algorithm for the classical mechanics with a Hubble expansion of the space included in the Newtonian dynamics
\begin{eqnarray}
\textbf{v}_k(t+\delta t/2)=\frac{(1+ \delta t H /2)\textbf{v}_k(t-\delta t/2) + \delta t/m_k\textbf{f}_k(t)}{1-\delta t H/2}, \nonumber \\
	\textbf{r}(t+\delta t)= \textbf{r}(t)+ \delta t  \textbf{v}(t+\delta t/2).
					 \end{eqnarray}
The discrete classical dynamics with Hubble expansion, Eqn. (52), is still time-reversible, but it  increases
the velocities, the momenta, and the angular momenta.

A galaxy, including the Milky Way, contains hundred of billion of stars, and a substantial amount of baryonic 
gas \cite{Gupta2012,Bergma2018} and
it is not possible directly to obtain the Newtonian dynamics of a galaxy with this number of objects. Instead, 
  models of small ''galaxies" of hundred of objects in
orbits around their center of gravity were simulated in \cite{Toxvaerd2022b}, and in an expanding universe with various strengths of the expansion.
 If these MD systems shall simulate the dynamics of a galaxy in the expanding universe, then one must relate $\textit{distances, times,}$  and $\textit{Hubble expansion}$ in the MD systems with
 the corresponding distances, times and Hubble expansion in a galaxy. Doing so
the value of the
Hubble constant in the MD units   for the models in \cite{Toxvaerd2022b} of  the  Milky Way in the expanding universe  was 
$H= 5.\times 10^{-8\pm 1}$ in MD units.

Models of galaxies with $H= 0,$ $ 5. \times 10^{-8},$ $ 5.  \times 10^{-7}$ and $5.  \times 10^{-6}$,
respectively, were simulated over a very long time with many billion of time steps and 
corresponding to more than 25 billion years in cosmological time. Figure 8 shows the number of objects in the central part of the   galaxies with a mean distance
  $\bar{r}(t) < 15000 \approx$ 15 kilo parsec to the center of gravity of the MD system as a function of time.
The galaxies with  with $H \leq 5.\times 10^{-8}$ 
contained twice as many bound objects in the ''halos" with distances $ 15000 < \bar{r}(t) < 100 000$.
The models of the Milky Way with $H \leq 5.\times 10^{-8}$ were 
rather stable even for times which corresponds to more than  twice the age of the Universe in contrast to
 galaxies with $ H >  5.\times 10^{-8}$. The release of the last bound object in
a galaxy with  $ H =  5.\times 10^{-7}$ is shown in Figure 9. The last bound object  escaped the gravitational center, but first  after
 $ t=  7.5 \times 10^{6}$ (3 billion MD time steps) or	what corresponds to $\approx$ 13 billion years after the galaxy was created.  
The rotating galaxies with  $H \leq 5.\times 10^{-8}$
released an object from time to time, but they contained still many bound objects at the end of the simulations corresponding to more than 25 Gyr (Figure 8).

 \begin{figure}
\begin{center}
\includegraphics[width=5cm,angle=-90]{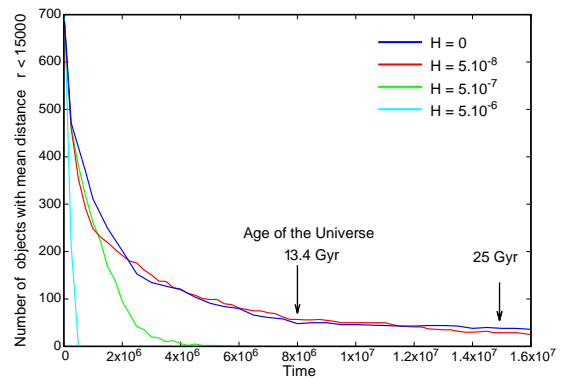}
	\caption{ Number of objects in the ''disk'' in the galaxies with mean distances $\bar{r}(t) < 15000$ to the mass centers 
	of the galaxies as a function of time. The systems
were exposed to different strengths of Hubble expansions given in the figure.}
\end{center}
\end{figure}

 \begin{figure}
\begin{center}
\includegraphics[width=6cm,angle=-90]{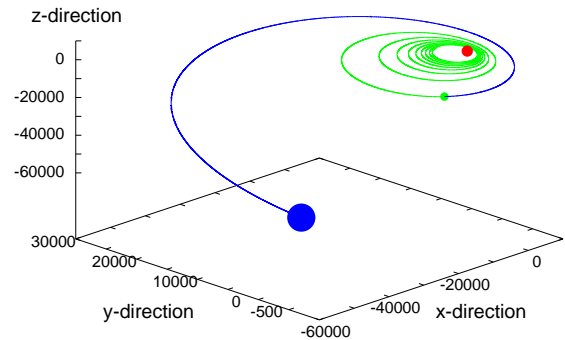}
	\caption{  The last bound object in an unstable galaxy in a fast expanding universe with $H= 5.\times 10^{-7}$. The object escaped the gravitational center (enlarged red sphere)  after
	what corresponds to $\approx$ 13 billion years after the galaxy was created. The first part of
	its dynamics ($0 < t < 6.5 \times 10^6 \approx 10.9 $ Gyr) is shown with green, and the escape from the center of gravity is shown with blue. 
	The galaxies are, however, stable for a Hubble expansion ten times weaker
	and equal to the Hubble expansion of our
	Universe. }
\end{center}
\end{figure}

The dynamics of galaxies in an expanding universe are often determined by gravitational  and dark matter in an Einstein-de Sitter universe \cite{Vogelsberger2019},
or alternatively by modifying the gravitational long-range attractions in the Newtonian dynamics (MOND) \cite{Milgrom1983}.
The Milky Way has, however, only performed $\approx$ 60 rotations after its creation and the galaxy is hardly in any kind of a steady state, and
the simulations in \cite{Toxvaerd2022b} with pure gravitational forces indicate, that the explanation for the dynamics of galaxies may be that the Universe is very young in cosmological times.
Although the models of the Milky Way release objects from time to time they still contained many bound objects at 25 Gyr, which
is almost twice the age of
the Universe.  The Hubble expansion will, however, sooner or later release the objects in the galaxies, but
the simulations indicate that this will first happen in a faraway future.

\section{Conclusion}

Computer simulation of the time evolution in a complex classical system, Molecular Dynamics (MD), is a standard method,
and it is used in numerous scientific articles in Natural Science. Newton's discrete algorithm, Eq. (3) is used
 in almost all the simulations,
albeit it is not acknowledged nor known that it was Newton who proposed it at the beginning of his book $Principia$ (Section 2).
Usually, the computational algorithm in the MD simulations is even not mentioned in the articles. MD is a standard tool and the algorithm is a ''black box". MD with
Newton's discrete algorithm from $Proposition$ $I$ is, however, exact in the same sense as Newton's analytic counterpart, the Classical Mechanics.
The discrete dynamics is time reversible, symplectic, and has the same invariances as the analytic dynamics (Section 3.1).

There is, however, no qualitative difference between the two dynamics. This is due to the fact, that there exists a ''shadow Hamiltonian''
 nearby the Hamiltonian for the analytic dynamics. The shadow Hamiltonian can be obtained by an asymptotic expansion, and
the positions generated by the discrete Newtonian dynamics are located on the analytic trajectories for the shadow Hamiltonian (Section 3.2).

It is only possible to obtain the solution of  Newton's classical differential equations for a few simple systems, e.g. for a harmonic oscillator. But
the discrete Newtonian dynamics can be obtained for almost all classical 
systems without any problems, e.g. for complex celestial systems (Section 4.1, 4.2, and 4.3). 

The fact that there exist two equally valid formulations of classical dynamics, the discrete Newtonian dynamics and the analytic Classical Mechanics raises the question:
What is the classical limit of quantum mechanics?  Classical Mechanics and analytic quantum mechanics  are connected by the Wigner expansion \cite{Wigner},
and Lee and coworkers have formulated a discrete nonrelativistic quantum mechanics \cite{Lee1,Lee2,Lee3}, where Newton's discrete dynamics is
the classical limit (Section 3.3).
 The difference in the energy between the
 analytic energy and the energy obtained by Eq. (20) for   a  time quant $t_p$  is of the
order $t_p^2$. With $t_p$ equal to Plank's time quant the difference is absolute marginal. The Heisenberg uncertainty between positions and momenta is of the order $t_p$  and
this uncertainty is
an inherent quality of discrete dynamics.
But the analytic  quantum electrodynamics (QED)   is in all manner  fully appropriate and
there is a lack of justification for preferring discrete quantum mechanics.

				\section{Acknowledgment}
 Niccol\`{o} Guicciardini, Ole J. Heilmann, Isaac Newton  and Ulf R. Pedersen are gratefully acknowledged.
This work was supported by the VILLUM Foundation’s Matter project, grant No. 16515.

\appendix
\section{Verlet algorithm}

The first 
Molecular Dynamics simulation of   systems of particles with  analytic  potentials was published in 1964 by A. Rahman \cite{Rahman1964}, where he 
used a higher-order predictor-corrector algorithm for the determination of the new positions $\textbf{r}^N(t+ \delta t)$ from the previous discrete positions.
Loup Verlet (1931-2019) published in 1967 the article \textit{ Computer ''Experiments" on Classical Fluids. I. Themodynamical
Properties of Lennard-Jones Molecules} \cite{Verlet1967}, where his algorithm, Eq (4),
\begin{equation}
	\textbf{r}(t+\delta t)= 2 \textbf{r}(t)-\textbf{r}(t-\delta t) +\frac{\delta t^2}{m} \textbf{f}(t)
\end{equation}	
was introduced without any explanation. Loup Verlet  was in the mid-nineteen hundred and sixties affiliated 
with J. L. Lebowith at Yeshiva University, NY.
 Lebowith gave a preliminary report of Verlet's simulation at a conference in Copenhagen in 1966, where I
became acquainted with the Verlet algorithm and Discrete Molecular Dynamics.
According to my supervisor at Copenhagen University Eigil L. Pr\ae stgaard,
who   was a postdoc at Yeshiva University in the same period, the algorithm was derived by
a forward and backward Taylor expansion
\begin{eqnarray}
\textbf{r}(t+\delta t)= \textbf{r}(t)+ \delta t \frac{\partial \textbf{r}(t)}{\partial t}
+\frac{1}{2} \delta t^2 \frac{\partial^2 \textbf{r}(t)}{\partial t^2}+
	\frac{1}{6}\delta t^3\frac{\partial^3 \textbf{r}(t)}{\partial t^3}+ \mathcal{O}(\delta t^4) \nonumber 	\\
\textbf{r}(t-\delta t)= \textbf{r}(t)- \delta t \frac{\partial \textbf{r}(t)}{\partial t}
+\frac{1}{2} \delta t^2 \frac{\partial^2 \textbf{r}(t)}{\partial t^2}-
	\frac{1}{6}\delta t^3\frac{\partial^3 \textbf{r}(t)}{\partial t^3}+ \mathcal{O}(\delta t^4),
\end{eqnarray}
and the algorithm was obtained from  the sum  $ \textbf{r}(t+\delta t)+\textbf{r}(t-\delta t)$ and
$ \delta t^2 \partial^2 \textbf{r}(t)/\partial t^2=\frac{\delta t^2}{m} \textbf{f}(t)$. All the odd  terms in A.2 cancel, and the Verlet algorithm is a 
four-order time symmetric predictor of the positions at the analytic trajectories. The scientific community and Verlet were
first much later aware, that it
actually was Newton who first published the geometric formulation of the algorithm in $Proposition$ $I$ \cite{Nauenberg2018a}.

 Today almost all MD simulations in physics and chemistry are performed with the algorithm, which appears under a variety of different names.


\begin{thebibliography}{99}
\bibitem{Newton1687} I. Newton, PHILOSOPHI\AE \ NATURALIS PRINCIPIA MATHEMATICA. \textit{LONDINI}, \textit{Anno} MDCLXXXVII.
\bibitem{Verlet1967} L. Verlet, Phys. Rev.  {\bf 159}, 98 (1967).
\bibitem{Tildesley} M. P. Allen,  D. J. Tildesley,  {\it Computer Simulation of Liquids}
 (Oxford Science Publications, Oxford, 1987).
\bibitem{Frenkel} D. Frenkel, B. Smit,  {\it Understanding Molecular Simulation} (Academic, New York, 2002).	
\bibitem{toxone}  S. Toxvaerd, Phys Rev. E, {\bf 50}, 2271 (1994).  The word $\textit{shadow Hamiltonian}$
 was introduced in this paper, inspired by the terms
 $\textit{a slightly perturbed Hamiltonian}$ [H. Yoshida,  Phys. Lett. A   {\bf 150}, 262 (1990)]
  and $\textit{ shadow trajectories}$  [C. Grebogi, S. M. Hammel, J. A: Yorke and T. Saur, 
  Phys. Rev. Lett.   {\bf 65}, 1527 (1990)].
\bibitem{Newtonengtrans}  I. B. Cohen, A. Whitman, J. Budenz, {\it The Principia : Mathematical Principles of Natural Philosophy  } (Univ. California Press, Berkeley 1999).
\bibitem{Toxvaerd2020} S. Toxvaerd, Eur. Phys. J. Plus   {\bf 135}, 267 (2020).
\bibitem{Chandrasekbar} S. Chandrasekbar,    {\it Newon's Principia for Common Reader} (Clarendon Press, Oxford, 1995) Page 76-77.
\bibitem{Nauenberg2019} M. Nauenberg,   Ann. Sci.     {\bf 76}, 1  (2019).
\bibitem{Erlichson1997} H. Erlichson,  Historia Mathematica   {\bf 24}, 167  (1997).
\bibitem{Nauenberg1998} M.  Naueberg,   Historia Mathematica   {\bf 25}, 89  (1998).
\bibitem{Nauenberg2005} M. Naueberg,  Phys. in Perspec.    {\bf 226}, 1  (2005).
\bibitem{Pourciau2011} B. Pourciau,   Am. J. Phys.     {\bf 79}, 1015  (2011).
\bibitem{Nauenberg2012} M. Nauenberg,   Am. J. Phys.     {\bf 80}, 931  (2012).
\bibitem{Pourciau2012} B. Pourciau,   Am. J. Phys.     {\bf 80}, 934  (2012).
	\bibitem{Guicciardini1999} N. Guicciardini, 
{\it Reading the Principia. The debate on Newton’s mathematical methods for natural philosophy from 1687 to 1736},
		(Cambridge University Press. 1999).
\bibitem{Nauenberg2014} M. Naueberg, Arch. Hist. Exact Sci.   {\bf 68}, 179 (2014).
\bibitem{Coelho2018} R. L. Coelho,  Acta Mech.  {\bf 229}, 2287 (2018).
\bibitem{Friedman1991} A. Friedman, S. P. J. Auerbach,  J. Comput. Phys.  {\bf 93}, 177 (1991), $ibid$  {\bf 93}, 189 (1991).
\bibitem{Goldstein}  H. Goldstein, {\it Classical Mechanics},( Addison-Wesley Press Second Ed. 1980), Chap. 1. 
\bibitem{Toxvaerd2014} 	 S. Toxvaerd, J. Chem. Phys.  {\bf 140}, 044102 (2014).
\bibitem{Griffiths} D. F. Griffiths, J. M. Sanz-Serna, SIAM J. Sci. Stat. Comput. {\bf 7}, 994 (1986).
\bibitem{Sanz-Serna} J. M. Sanz-Serna, Acta Numer. {\bf 1}, 243 (1992).
\bibitem{Hairer} E. Hairer, Ann. Numer. Math. {\bf 1}, 107 (1994).
\bibitem{Reich} S. Reich, SIAM J. Numer. Anal.  {\bf 36}, 1549 (1999).
\bibitem{Gans} J. Gans, D. Shalloway, Phys. Rev. E  {\bf 61}, 4587 (2000).
\bibitem{toxtwo}  S. Toxvaerd, O. J. Heilmann, J. C. Dyre, J. Chem. Phys.  {\bf 136}, 224106 (20012).
\bibitem{MDunits}  In MD the mass $m$ is usually included in the time unit.
 The unit length, energy and time  for LJ systems are, respectively,
		$\sigma$, $\epsilon$ and $\sigma\sqrt{m/\epsilon}$. Temperature is $k_{\textrm{B}}T/\epsilon$.
For MD details, see  S. Toxvaerd, Mol. Phys. {\bf 72}, 159 (1991).
\bibitem{toxa}  S. Toxvaerd.  J. Chem. Phys. {\bf 137}, 214102 (2012).
\bibitem{Lagrange1788} Lagrange  M\'{e}canique analytique 1788-1789.
\bibitem{Wigner}  E. Wigner, Phys. Rev. {\bf 40}, 749 (1932).
\bibitem{Lee1} T. D. Lee, Phys. Lett.  {\bf 122 B}, 217 (1983).
\bibitem{Lee2} T. D. Lee, J. Stat. Phys.  {\bf 46}, 843 (1987).
\bibitem{Lee3}  R. Friedberg,  T. D. Lee, Nucl. Phys.   {\bf B 225 [FS9]}, 1 (1983).
\bibitem{Garay}  L. J. Garay, J. Mod. Phys. A  {\bf 10}, 145 (1995).
\bibitem{Regge2000} Regge, T., Williams, R. M. , J. Math. Phys.   {\bf 41}, 3964  (2000).
\bibitem{Toxvaerd1991}   Toxvaerd, S., Mol. Phys.   {\bf 72} 159 (1991).
\bibitem{Levesque1993} D. Levesque, L. Verlet, J. Stat. Phys.  {\bf 72} 519 (1993).
\bibitem{Toxvaerd2011} S.  Toxvaerd, J. C. Dyre, J.Chem. Phys.  {\bf 134} 081102 (2011).
\bibitem{Newtonshell}  $Principia$, THEOREM XXXI.
\bibitem{Toxvaerd2022} S.  Toxvaerd, Eur. Phys. J. Plus  {\bf 137} 99 (2022).
\bibitem{Toxvaerd2022a}  S. Toxvaerd, Celest. Mech. Dyn.  {\bf 134} 40 (2022).
\bibitem{Aoki1992} S. Aoki,  Arch. Hist. Exact Sci. {\bf 44}, 147 (1992)
\bibitem{Vogelsberger2019} M. Vogelsberger, et al., NATURE REVIEWS PHYSICS   {\bf 2} 42 (2019).
\bibitem{Milgrom1983} M. Milgrom, M., 1983, Ap. J.  {\bf 270}, 371 (1983). 
\bibitem{Toxvaerd2022b}   Toxvaerd, S., CQD   {\bf 39}, 225006 (2022).
\bibitem{Hubble1929} E. Hubble, Proc. Nath. Acad. Sci.   {\bf 15}, 168 (1929).
\bibitem{Lan2022} T.-J. Zhang, C. Ma, T. Lan, Adv. Astron. {\bf 2010}, 184284 (2010).
\bibitem{Book} B. W. Carrol, D. A. Ostlie,  \textit{An Introduction to Modern Astrophysics}(Addiso-Wesley Publishing Company, Reading 1996)
\bibitem{Riess1998}  A. G. Riess, A. V. Filippenko, P. Challis,  et al. 1998, Astronomical J. 	 {\bf 116}, 1009 (1998).
\bibitem{Perlmutter1999} Perlmutter, S, Aldering, A., Goldhaber,  et al., Astrophys. J. {\bf 517}, 565 (1999).	
\bibitem{Soltis2021}J. Soltis, S. Casertano, A. G. Riess,  2021, Astrophys. J. Lett.  {\bf 908}: L5 (2021). 
\bibitem{Gupta2012} A. Gupta, S. Mathur,  Y. Krongold, F. Nicastro, M.  Galeazzi, Astrophys. J. Lett. {\bf 756}:L8 (2012). 
\bibitem{Bergma2018}  J. N. Bergma, M. E. Anderson, J. M. Miller, E. Hodges-Kluck, X. Dai,
	J.-T.	Li, Y. Li,  Z. Qu, Astrophys. J. {\bf 862}:3 (2018).
\bibitem{Rahman1964} A. Rahman,  Phys. Rev.  {\bf 136}, A405 (1964).
\bibitem{Nauenberg2018a}  M. Nauenberg, Am. J. Phys. {\bf 86}, 765 see Reference No. 7. (2018).						\
\end{thebibliography}
\end{document}